\documentclass[a4paper,11pt]{article}
\usepackage{jcappub} 

\usepackage{orcidlink}

\newcommand{\Lagr}{\mathcal{L}}

\title{\boldmath Inflationary dynamics of non-minimally coupled $f(R)$ matter-curvature theories}

\author[a,b]{Miguel Barroso Varela \orcidlink{0009-0006-9844-7661}}
\affiliation[a]{Departamento de Física e Astronomia, Faculdade de Ciências, Universidade do Porto, Rua do Campo Alegre s/n, 4169-007 Porto, Portugal}
\author[a,b]{Orfeu Bertolami \orcidlink{0000-0002-7672-0560}}%
\author[b,c]{and Andreas Mantziris \orcidlink{0000-0001-5899-670X}}
\affiliation[b]{Centro de Física das Universidades do Minho e do Porto, Rua do Campo Alegre s/n, 4169-007 Porto, Portugal}%
\affiliation[c]{Faculdade de Ci\^encias e Tecnologia da Universidade de Coimbra, Rua Larga, 3004-516 Coimbra, Portugal}

\emailAdd{up201907272@edu.fc.up.pt}
\emailAdd{orfeu.bertolami@fc.up.pt}
\emailAdd{andreas.mantziris@fc.up.pt}

\abstract{This study examines how inflationary dynamics are affected by $f(R)$-theories with a non-minimal coupling between matter and curvature. Both positive and negative corrections to the minimal coupling of General Relativity are considered, and a robust numerical method is developed that evolves the metric and the inflaton field in this modified theory beyond slow-roll. Through a stability analysis, we find that positive models are inherently unstable during slow-roll, whereas negative ones can accommodate a stable attractor de Sitter solution. Using the amplitude of the scalar power spectrum from the latest data releases, we constrain the scale of the non-minimal coupling to be above $10^{13}$ GeV. In light of the 2018 Planck, BICEP/Keck and the recent Atacama Cosmology Telescope data for the scalar spectral index and tensor-to-scalar ratio, strong constraints on the coupling strength force the effects of these modified theories to be, at most, slightly above the perturbative level. Furthermore, we determine that the choice of the perfect fluid matter Lagrangian does not impact the inflationary observables at the pivot scale. Finally, we present the predicted observables for different inflationary potentials and show that even though classical gravity is still preferred by the data, there are areas of the parameter space that are viable for non-minimally coupled inflationary models.}

\begin{document}
\maketitle
\flushbottom

\section{Introduction}\label{sec:Introduction}

The leading paradigm for the description of the early universe is cosmological inflation, a period of accelerated expansion due to a scalar field slowly rolling down an approximately flat potential \cite{Guth:1980zm, Starobinsky:1979ty, Starobinsky:1980te}. At the end of inflation, the inflaton field decays into Standard Model (SM) particles in a thermal process that (re)heats the universe into the Hot Big Bang (HBB) epoch. Although inflation addresses some of the long-standing problems of HBB cosmology, it also provides the mechanism through which primordial quantum fluctuations in the matter density field seed the large-scale structure of the late universe \cite{Liddle:2000cg}. The properties of these fluctuations are consistent with a multitude of observational probes, the most remarkable of which are the Cosmic Microwave Background (CMB) anisotropies \cite{Planck:2019evm}.

The addition of the inflaton field is usually justified from a particle physics perspective as a minimal scalar extension Beyond the SM (BSM) \cite{Guth:1980zm}, but gravitational origins can also be considered, as in the Starobinsky model of inflation \cite{Kofman:1985aw} for example. Furthermore, different metric formulations of gravity, such as Brans-Dicke theory \cite{Brans:1961sx} for instance, could allow the inflationary phase transition {of ``old'' inflation} to be completed via bubble nucleation \cite{La:1989st}, without fine-tuning the effective potential to ensure slow roll as in General Relativity (GR) \cite{Guth:1980zm}. The difficulty in building inflationary models in the context of string theory \cite{Kachru:2003sx}, which at lowest order in the Regge slope parameter ($\alpha'$) power expansion leads to GR, suggests that alternative theories of gravity could offer a viable alternative. In fact, the recent discussion of the conditions to achieve suitable effective theories arising from string theory, the so-called swampland conjectures \cite{Vafa:2005ui, Ooguri:2006in,Ooguri:2018wrx,Obied:2018sgi}, are shown to be incompatible with slow-roll conditions for single-field inflation (see, for instance, Refs. \cite{Kehagias:2018uem, Kinney:2018nny}). However, slow-roll conditions satisfy the swampland conjectures in the context of warm inflation \cite{Brandenberger:2020oav, Bertolami:2022knd} and Chaplygin-inspired inflation \cite{Bertolami:2024xkw}. In this work, we adopt a ``hybrid'' approach, where we consider a BSM inflaton that is non-minimally coupled (NMC) to curvature with higher-order geometrical terms. In this context, no match of the swampland conditions can be achieved with a positive $R^3$ term \cite{Bertolami:2023bih}, but even though we study both positive and negative power-law terms here, the NMC effects being subdominant implies that the same conclusion can be extended to include general negative $R^n$ terms. These considerations have the obvious merit of narrowing the range of cosmologically interesting models that are worthy of UV embeddings to string theory \cite{Berera:2023mlj, Raveri:2018ddi, Garg:2018reu}, which nevertheless corresponds to just one, although favourable, approach to quantum gravity. Therefore, whether these criteria can decisively determine the viable theories that describe the early universe is not conclusive at this moment \cite{Kinney:2018nny, Denef:2018etk, Casse:2022ymj}.

Our approach is motivated by a class of models called $f(R)$ theories \cite{Buchdahl:1970ynr,Sotiriou:2008rp, DeFelice:2010aj,Capozziello:2011et}, where a natural extension of GR is achieved by including higher-order curvature correction terms to the Einstein-Hilbert term of GR. As mentioned previously, we shall consider the NMC matter-curvature model, which was originally introduced in Refs. \cite{Gonner:1976gq, Gonner:1984zx} and studied as a source of an extra force in the geodesic equation to address the abnormal galaxy rotation curves in Refs. \cite{Bertolami:2007gv,Bertolami:2011ye}. The NMC models have been extensively researched as a gravitational mechanism for the late-time accelerated expansion of the universe \cite{Bertolami:2010cw}, while also modifying the standard evolution of cosmological parameters from the CMB that restores their compatibility with direct measurements from type Ia supernovae and removes the Hubble tension \cite{BarrosoVarela:2024htf,BarrosoVarela:2024ozs}. Furthermore, its implications on density perturbations and structure growth have been examined in Ref.\cite{Bertolami:2013kca,BarrosoVarela:2025mro}, whereas predictions on metric perturbations, along with propagation of the tensorial gravitational wave modes and the presence of additional scalar polarizations were discussed in Refs. \cite{Bertolami:2017svl,BarrosoVarela:2024egg}. Most importantly for this work, its effect on inflationary properties was determined in Ref. \cite{Gomes:2016cwj} for a set of power-law models, placing the modified theory within observational constraints.  

The significance of considering NMC theories in the early universe can also be viewed from the perspective of Quantum Field Theory on curved spacetime. In the simplest case, non-minimal coupling terms between fields and gravity are radiatively generated and necessary for the renormalizability of the theory \cite{Birrell:1982ix}. Furthermore, given the high energy scales of the primordial universe, such couplings cannot be tuned to zero universally, since their running with renormalization scale will render them finite at some other energy scale \cite{Mantziris:2020rzh}. In most scenarios regarding this early epoch, some aspects of NMC matter to curvature not only need to be considered, but can lead to qualitatively different behaviour, such as the stabilisation of the electroweak vacuum \cite{Herranen:2014cua, Herranen:2015ima, Shaposhnikov:2020geh, Figueroa:2017slm, Markkanen:2018bfx, Markkanen:2018pdo, Mantziris:2020rzh, Vicentini:2020lhm, Li:2022ugn, Mantziris:2022fuu, Glavan:2023lvw, Laverda:2024qjt, Laverda:2025pmg}, curvature-induced phase transitions \cite{Bettoni:2019dcw,Bettoni:2021zhq, Laverda:2023uqv, Kierkla:2023uzo, Goertz:2024gzw, Mantziris:2024uzz, Aldabergenov:2025oys, Bettoni:2024ixe, Rubio:2025egw}, contributions to the potential of the inflaton \cite{Rubio:2018ogq, McDonough:2020gmn, Lebedev:2021xey, Figueroa:2021iwm, Cheng:2022pzs, Barman:2023opy, Poisson:2023tja, Piani:2025dpy, Rigouzzo:2025hza, Ahmed:2025rrg, Kersten:2024ucm}, quintessence \cite{Perrotta:1999am, Bartolo:1999sq, Almatwi:2023cir, Carloni:2024ybx, Irges:2025idm}, or scalar dark matter \cite{Cosme:2018wfh, Li:2021fao, Clery:2022wib, Silveravalle:2025yij}. These considerations, in conjunction with the gravitational narrative mentioned above, have motivated us to pursue this in-depth study of the dynamical behaviour of an NMC inflaton within the context of $f(R)$ theories.

This paper is organised as follows. We present the framework of NMC models along with the field and conservation equations and their associated implications on homogeneous cosmological backgrounds in Section \ref{sec:NMCModel}. In Section \ref{sec:NMCInflation} we study the compatibility of power-law NMC models with exponential expansion in the slow-roll regime, compute the corresponding imprints of the theory on the inflationary observables, and describe the numerical method used for calculating the behaviour of the inflationary dynamics beyond slow-roll. In Section \ref{sec:Results}, we showcase the theoretical predictions for inflationary observables for various models of cold ``new'' inflation, and discuss the problem of graceful exit in ``old'' inflation in the NMC context. Finally, our conclusions are presented in Section \ref{sec:Conclusions}. Throughout this paper, we use the metric signature $(-,+,+,+)$ and the conventions $c=1$ for the speed of light and $\kappa^2=8\pi G=M_P^{-2}=1$ for the gravitational constant $G$ and the reduced Planck mass $M_P = 2.435 \times 10^{18}$ GeV. We denote the value of a quantity at the pivot scale with the subscript ``*'', during slow roll with ``inf'', and at the end of inflation with ``end''.

\section{Non-minimally coupled matter to higher-order gravity}\label{sec:NMCModel}
\subsection{Action and field equations}
The family of NMC matter-curvature $f(R)$ theories \cite{Bertolami:2007gv} is given by the action 
\begin{align}
    S=\int dx^4 \sqrt{-g}\left[\frac{1}{2}f_1(R)+f_2(R)\mathcal{L}_m\right] \, ,
    \label{eq:general_action}
\end{align}
where $f_i (R)$ are arbitrary functions of the curvature scalar $R$, $g$ is the determinant of the spacetime metric, and $\Lagr_m$ is the matter Lagrangian density, which is non-minimally coupled to curvature when $f_2(R)\neq1$. GR is recovered in the minimal coupling regime $f_2(R)=1$ when imposing the Einstein-Hilbert action with $f_1(R)=R$. A cosmological constant can be included in the minimally coupled sector via $f_1(R)=R-2\Lambda$. This action is defined differently in some related works, such as its definition in Ref. \cite{Bertolami:2007gv}, where the convention $f_2\rightarrow(1+f_2)$ is used.

Varying the action with respect to the metric $g_{\mu\nu}$ yields the field equations \cite{Bertolami:2007gv}
\begin{equation}\label{eq:FieldEquations}
    FG_{\mu\nu}=f_2T_{\mu\nu}+\Delta_{\mu\nu}F+\frac{1}{2}g_{\mu\nu}(f_1-F R) \, ,
\end{equation}
where we have defined $\Delta_{\mu\nu}\equiv\nabla_\mu\nabla_\nu-g_{\mu\nu}\Box \,$, $F_i=df_i/dR \,$, and
\begin{align}
    F = F_1 + 2 F_2 \Lagr_m \, .
    \label{eq:F}
\end{align}
The Einstein field equations $G_{\mu\nu}=T_{\mu\nu}$ are recovered in the GR limit, $f_1=R$ and $f_2=1$. Taking the divergence of the field equations and applying the Bianchi identities $\nabla_{\mu}G^{\mu\nu}=0$ we obtain the modified conservation equations \cite{Bertolami:2007gv}
\begin{equation}\label{eq:NonConservationEq}
    \nabla_\mu T^{\mu\nu}=\frac{F_2}{f_2}\left(g^{\mu\nu}\mathcal{L}_m-T^{\mu\nu}\right)\nabla_\mu R \, .
\end{equation}
This is explicitly dependent on $f_2$, which highlights that the non-conservation of the stress-energy tensor is due to the non-minimal coupling of the matter Lagrangian in the action. Eq. (\ref{eq:NonConservationEq}) naturally reduces to the standard conservation equation for $f_2=1$ and thus $F_2=0$. When describing a homogeneous and isotropic universe, we assume the matter content to be approximated by a perfect fluid with
\begin{equation}\label{eq:StressEnergyTensor}
    T^\mu_\nu=\text{diag}(-\rho,p,p,p) \, ,
\end{equation} 
where $\rho$ is the energy density and $p$ is the pressure.

A crucial property of this model is the explicit dependence of both sets of equations (\ref{eq:FieldEquations})-(\ref{eq:NonConservationEq}) on the {matter} Lagrangian density $\Lagr_m \, $, in contrast to entering only through the corresponding stress-energy tensor in GR. As originally shown in Ref. \cite{Brown:1992kc} and later discussed in the context of NMC theories in Ref. \cite{Bertolami:2008ab}, the perfect fluid Lagrangian choices of $\Lagr_m=p$ and $\Lagr_m=-\rho$ are degenerate in GR because they differ only by a surface term in the action. However, this is no longer the case in NMC models. During the slow-roll period, $p\approx-\rho$ and this degeneracy is approximately restored, as pointed out in Ref. \cite{Gomes:2016cwj}. Considering that the matter content during cold inflation is fully accounted for by the inflaton field $\phi$ rolling down its potential $V(\phi)\,$, the matter Lagrangian is given by 
\begin{align}
    \Lagr_m =(2\beta-1)\frac{\phi^2}{2}-V(\phi)  \quad \text{where} \quad\beta=
\begin{cases} 
     1\, , & \text{for } \mathcal{L}_m=p \, . \\
    0 \, ,  & \text{for } \mathcal{L}_m=-\rho \, .
\end{cases} \, \quad 
    \label{eq:Lagr_m}
\end{align}

\subsection{Cosmology in non-minimally coupled models}

An expanding homogeneous and isotropic universe is described by the Friedmann–Robertson-Walker (FRW) metric 
\begin{equation} \label{eq:FRW}
    ds^2=-dt^2+a^2(t)\sum_{i=1}^3dx_i^2 \, ,
\end{equation}
which is inserted into the modified field Eqs. (\ref{eq:FieldEquations}) to yield the modified Hubble rate,
\begin{equation}
     H^2 = \frac{1}{6F} \left[2f_2 \rho -6H\dot F +FR-f_1  \right] \, .
    \label{eq:GeneralFriedmannEq}
\end{equation}
This NMC Friedmann equation is now a second-order differential equation for $H$, as $R=6(2H^2+\dot H)$ and $\dot F\sim\dot R\sim\ddot H$. Additionally, the Raychaudhuri equation is given by
\begin{equation}
    -2F(2\dot H+3H^2)=2\ddot F+4H\dot F+f_1-FR+2f_2p \, ,
    \label{eq:GeneralRaychaudhuriEq}
\end{equation}
which has even higher-order derivatives of $H$ due to the $\ddot F$ term. We can rewrite both of these equations in terms of the pressure and density components of the stress-energy tensor as
\begin{align}
    2f_2\rho &= (6FH^2-FR+f_1)+6\dot F H \,,
    \label{eq:RhoFriedmann} \\
    2f_2p &= -(6FH^2-FR+f_1)-4\dot F H-4F\dot H-2\ddot F \, .
    \label{eq:PressureRaychauduri}
\end{align}
Evidently, the density and pressure of the inflaton obey $p=-\rho$ up to derivatives of $F$ and $H$, which vanish in the slow-roll regime where $H$ and $\rho$ are approximately constant.

Inserting the FRW metric (\ref{eq:FRW}) and the stress-energy tensor (\ref{eq:StressEnergyTensor}) into the non-conservation equation (\ref{eq:NonConservationEq}) leads to the modified continuity equation
\begin{align}
    \dot\rho+3H(\rho+p)+\Gamma_c(\rho+\Lagr_m)=0 \,,
\end{align}
where we have defined the NMC-induced friction term as 
\begin{equation}
    \Gamma_c {(R, \dot R)} =\frac{F_2}{f_2}\dot R \, .
    \label{eq:GammaFriction}
\end{equation}
This can also be written in terms of the inflaton field with $\rho=\dot\phi^2/2+V(\phi) \, $, as
\begin{equation}\label{eq:ModifiedConservationEq}
    \ddot\phi+3H\dot\phi+V'(\phi) =-\beta\Gamma_c\dot\phi \, .
\end{equation}
During inflation, the Ricci scalar decreases, $\dot R<0$, so the additional friction term acts for or against the Hubble friction depending on the sign of $F_2\,$, although its effect is small during slow-roll, since in the standard cold inflationary models $|\dot R| \ll1$. However, once the inflaton gains some kinetic energy and spacetime starts to deviate from de Sitter (dS), the contribution from $\Gamma_c$ will become increasingly more significant. Additionally, the emergence of the NMC friction term $\Gamma_c$ raises yet another important distinction between the choice of the matter Lagrangian. As already mentioned, it is irrelevant during slow-roll since $p\approx-\rho$, but once the inflaton rolls down the potential and gains more kinetic energy, the differences between the choices of $\Lagr_m$ differ by up to a factor of $-1$ during kination ($p\approx\rho$). In this work, we expand the dS study of Ref. \cite{Gomes:2016cwj}, by performing a full numerical analysis of the inflationary dynamics and the effects of the matter Lagrangian choice beyond slow-roll. The choice of $\Lagr_m$ has direct consequences on both the conservation Eq. (\ref{eq:ModifiedConservationEq}) and the Friedmann Eq. (\ref{eq:GeneralFriedmannEq}) through its explicit dependence on $F$, but we found that this has no tangible impact on the CMB inflationary observables. However, the discovery of any characteristic behaviour beyond slow-roll could lead to new considerations on the choice of $\Lagr_m$, this time stemming directly from the early Universe, in contrast to arguments focussing on late-time modifications of the NMC theory \cite{BarrosoVarela:2025mro}. For example, if a particular Lagrangian choice exhibits pathological behaviour during inflation, this could hint at deeper issues of that Lagrangian in the context of high-energy corrections from GR to the non-minimal coupling and guiding us towards the most physically stable UV completion of the theory.

\section{Inflation in non-minimally coupled matter-curvature theories}\label{sec:NMCInflation}
\subsection{Compatibility and stability of slow-roll in purely NMC models}\label{subsec:SlowRoll_NMC}

In general, the modified Friedmann Eq. (\ref{eq:GeneralFriedmannEq}) is a higher-order differential equation that can no longer be solved analytically for the expansion rate as in GR. However, for an approximately static background energy density, it can be simplified to give at least an approximate solution for the cosmological metric. The slow-roll regime offers such a setting, with a constant inflaton density leading to a standard exponential expansion where $\dot H=0$, which, as we shall discuss later, is a non-trivial assumption that is unfulfilled for several models. Nevertheless, we will continue to consider this solution preliminarily, as we later discuss how it is a stable state in models with negative corrections to the minimal matter-curvature coupling from GR. The modified Friedmann Eq. (\ref{eq:GeneralFriedmannEq}) in slow-roll simplifies to 
\begin{align}\label{eq:SlowRollFriedman}
    H^2 \simeq  \left(\frac{f_2 }{1 - 2 F_2 \Lagr}\right) \frac{\rho}{3 }  \, . 
\end{align}
For a general power-law form for the NCM term,
\begin{align}
    f_2^{(\pm)}(R) = 1 \pm \left(\frac{R}{M^2} \right)^n \, ,
    \label{eq:f2}
\end{align}
where $n\geq2$ and $M$ sets the scale of the NMC, we can invert Eq. (\ref{eq:SlowRollFriedman}) for $R=12H^2$ in terms of the density
\begin{align}\label{eq:H(rho)}
    \rho^{(\pm)} = \frac{3 H^2}{1 \mp \frac{(n - 2)}{2}\left( \frac{12 H^2}{M^2}\right)^n} \, .
\end{align}
Interestingly, for $n=2$ the NMC Friedmann equation is always the same as in GR regardless of the sign in $f_2$ \cite{Gomes:2016cwj} and thus the inflaton is effectively decoupled from the non-minimal coupling during slow-roll. However, as we approach the end of inflation, this is no longer true, and a numerical simulation of the full inflationary dynamics is needed to distinguish between the standard inflationary paradigm and this modified scenario. In fact, since models with $n=2$ should differ from GR only away from slow-roll, this property offers a convenient check of the numerical implementation of the model, since in contrast to the weak coupling regime ($M\gg1)$, there can be sizeable NMC effects, which should approximately vanish for $n=2$ regardless of the strength of the coupling. 

\begin{figure}[ht!]
    \centering
    \includegraphics[width=0.493\linewidth]{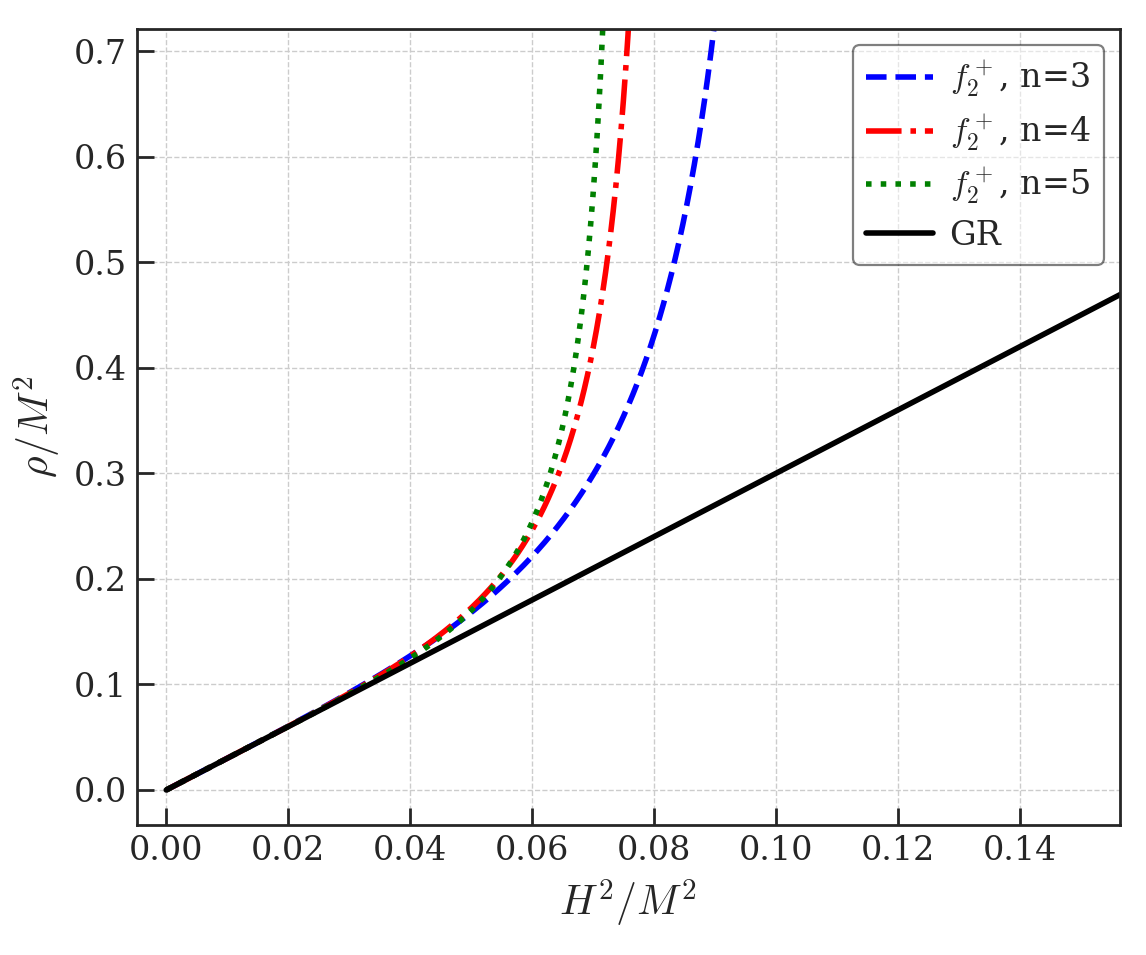}
    \includegraphics[width=0.4985\linewidth]{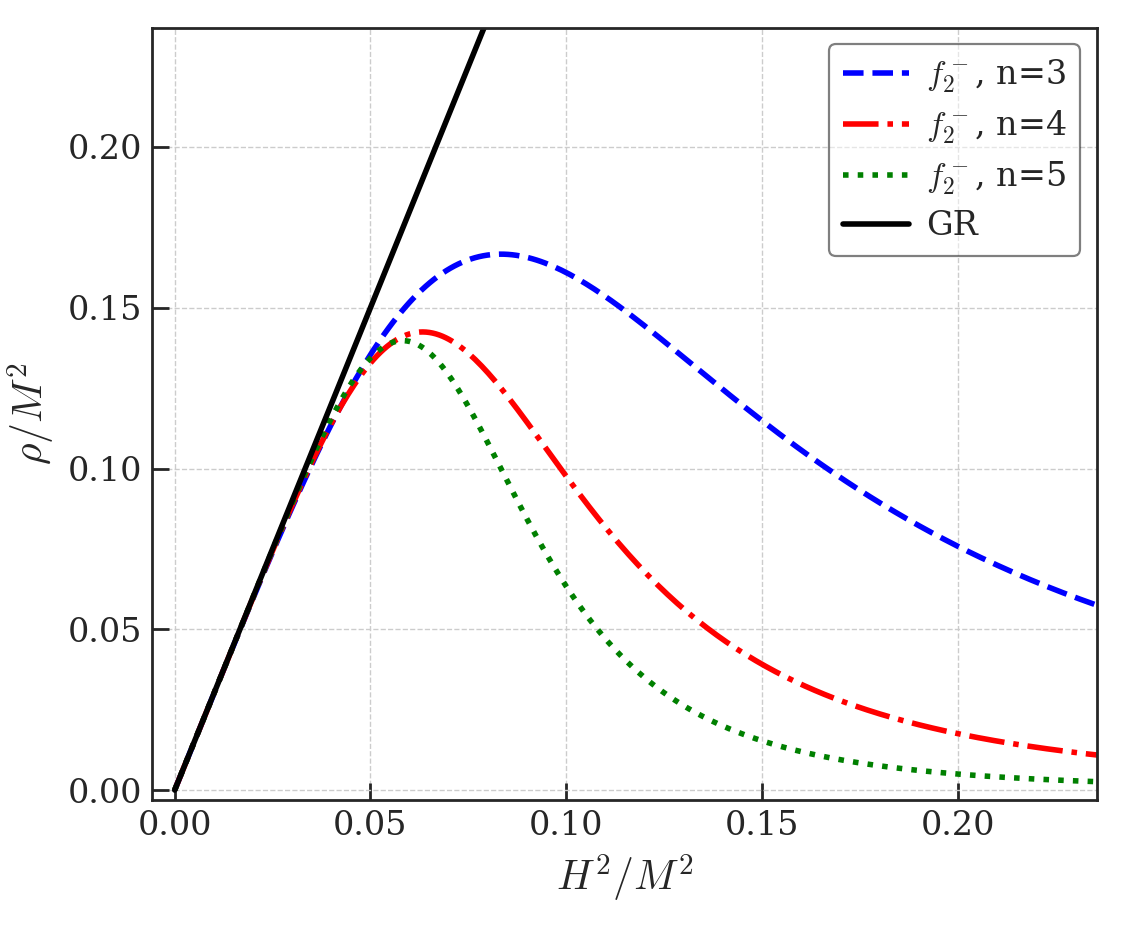}
    \caption{The slow-roll solutions for $\rho( H^2)$ under the $\dot H=0$ assumption in the $f_2^+$(left) and  $f_2^-$ (right) models. The GR solution $\rho=3H^2$ is shown for comparison.}
    \label{fig:FriedmannSolutionBranches}
\end{figure}

\begin{figure}[ht!]
    \centering
    \includegraphics[width=0.6\linewidth]{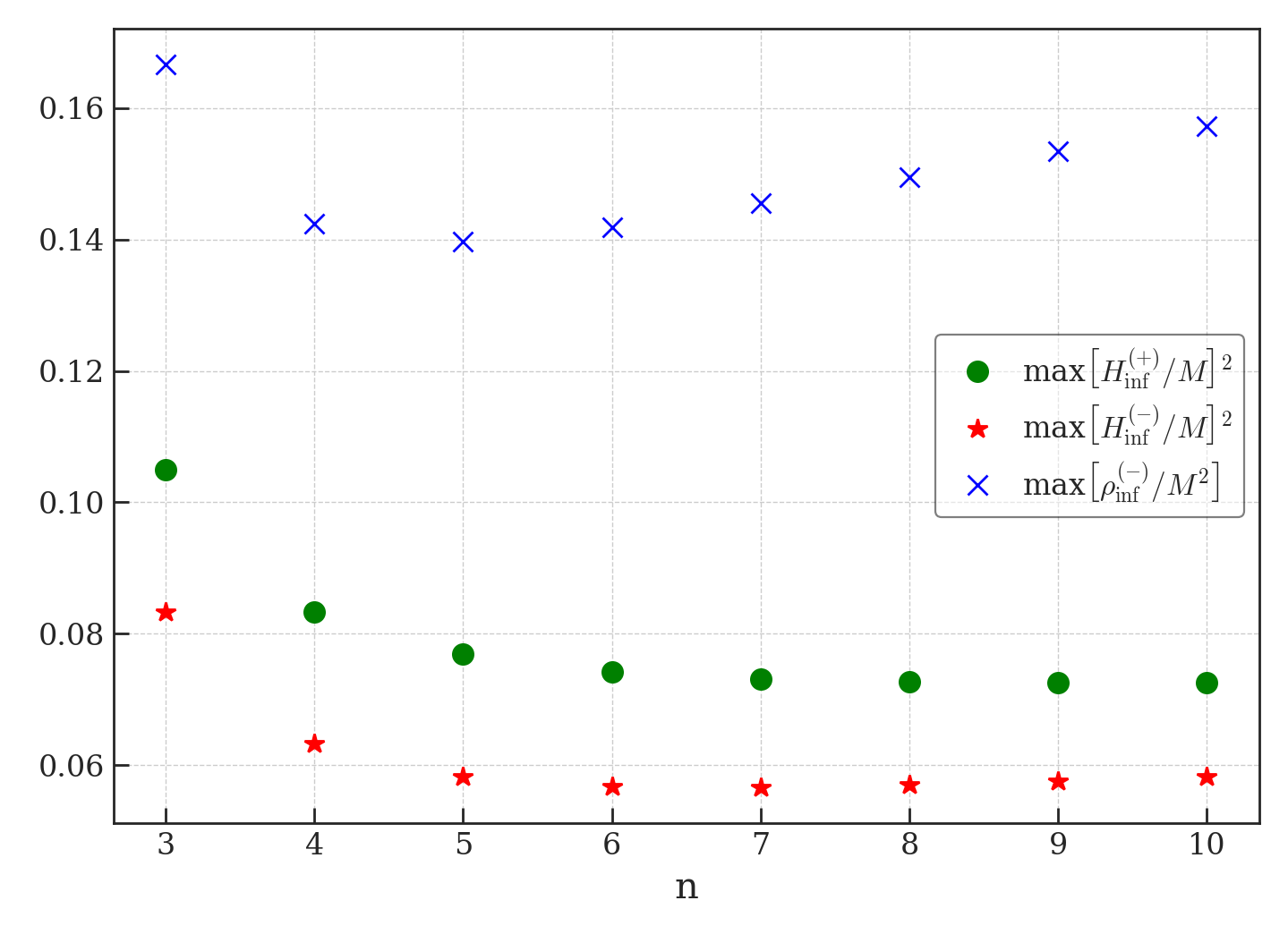}
    \caption{The maximum values of the inflationary expansion rate $H_{\text{inf}}$ and inflaton density $\rho_{\text{inf}}$ in the $(\pm)$-type models as functions of the power-law exponent $n$. The maximum density is unconstrained in $(+)$-type models and the corresponding maximum Hubble parameter is achieved asymptotically for large density values.}
    \label{fig:MaxHubble&Density}
\end{figure}

The functional dependency of $\rho(H^2)$ is shown in Figure \ref{fig:FriedmannSolutionBranches}, where it is evident that there are clear distinctions between the $(\pm)$-type models. For the positive-sign cases, we have a clear high-density limit that sets an upper bound on the inflationary scale,
\begin{equation}\label{eq:HighDensityLimit_Plus}
\max\left[H_{\rm inf}^{(+)}\right]^2=\lim_{\rho \rightarrow\infty}\left[ H^{(+)}_{\rm inf}\right]^2=\frac{M^2}{{12}}\left(\frac{2}{n-2}\right)^{1/n} \, .
\end{equation}
Even though Eq. (\ref{eq:H(rho)}) for $(-)$-type models is not injective for all values of $H_{\rm inf}^2>0$, there is an apparent physical branch of the solution, where the Hubble scale follows the decreasing energy density and tends asymptotically to GR, as shown in Fig. \ref{fig:FriedmannSolutionBranches}. This branch corresponds to inflationary densities and Hubble rates below the following upper bounds, 
\begin{align}
    \max\left[H_{\rm inf}^{(-)}\right]^2 &= \frac{M^2}{12}\left[\frac{2}{(n-1)(n-2)}\right]^{1/n} \, , 
    \label{eq:MaxHubble_Minus} \\
    \max\left[\rho_{\rm inf}^{(-)}\right] &= \frac{M^2(n-1)}{4n}\left[\frac{2}{(n-1)(n-2)}\right]^{1/n} \, ,
    \label{eq:MaxRho_Minus}
\end{align}
which are shown in Figure \ref{fig:MaxHubble&Density}. For $n \geq 2$, the $(+)$-type models have larger upper bounds on the inflationary scale than the $(-)$-type ones by a factor of $(n-1)^{1/n}$, although the maximum values of $H_{\text{inf}}$ in the latter are achieved at much lower densities. The determination of these upper bounds can help constrain the NMC coupling strength $M$ from realistic inflationary scales. For considerably larger and rather unphysical values of $n$, the NMC effects from both positive and negative prescriptions dominate and impose the characteristic NMC expansion timescale $H_{\rm inf} = 2^{\frac{1}{n}}M^2/12$ as the maximum Hubble rate.

In GR, the slow roll of the inflaton down its potential implies an exponentially expanding universe with a constant Hubble rate. However, the higher-order nature of the field equations in the NMC $f(R)$-theory means that this is no longer necessarily a valid approach, as the direct algebraic relation from GR ($H^2=\rho/3$) is replaced by the third-order differential Eq. \eqref{eq:GeneralFriedmannEq} for the expansion rate. During slow-roll, the inflaton Lagrangian is approximately the negative of its potential, $\Lagr_m = p \simeq - \rho \approx - V(\phi)$, which causes $(\rho+\Lagr_m)$ to vanish regardless of the Lagrangian choice in Eq. (\ref{eq:NonConservationEq}) and recovering GRr, $\dot\rho=0$. We may thus confidently assume that $\dot\Lagr_m=0$ during the slow-roll phase. By comparing Eqs. (\ref{eq:RhoFriedmann}) and (\ref{eq:PressureRaychauduri}), slow-roll implies that
\begin{equation}\label{eq:SlowRollDifferentialEq}
     \ddot F + 2F \dot H - \dot F H = 0 \, ,
\end{equation}
which is a second-order differential equation in $R$ and therefore a third-order differential equation in $H$,

\begin{multline}
    \dddot H =  - 4 \left( \dot H^2 + 4 H \ddot H \right) + H (4 H \dot H + \ddot H ) - \frac{(n-2) (4 H \dot H + \ddot H)^2}{2 H^2+\dot H} - \frac{2 \dot H (2 H^2 + \dot H )}{n-1} \\ 
    - \frac{6^{(1-n)} M^{2n} \dot H}{n (n-1) \left(2 H^2+\dot H \right)^{(n-2)} \Lagr_m} \, . 
\end{multline}
Unsurprisingly, it can be solved by an exponentially expanding universe with $\dot H=0$, but it is not manifestly evident that this is an attractor solution of the system. The above equation may be cast as a system of first-order differential equations in terms of the variables $u_i=\{H,\dot H,\ddot H\}$ where
\begin{equation}
\begin{cases}
 \dot u_1=u_2 \, ,\\
 \dot u_2=u_3 \, ,\\
 \dot u_3=\dddot H(u_1,u_2,u_3) \, ,
\end{cases}
\label{eq:system}
\end{equation}
such that we can analyse the stability of its solutions, specifically for the one responsible for exponential expansion with $u_i=\{H_{\rm inf},0,0\}$. According to the Jacobian and eigenvalues of the system, a solution is unstable if any of its eigenvalues is positive. We can check this for several values of $H_{\rm inf}$ and $M$, since $\Lagr_m=-\rho_{\rm inf}$ is uniquely determined in terms of these quantities by Eq. (\ref{eq:H(rho)}). For each $M$, we can determine the maximum allowed value of $H_{\rm inf}$ from Eqs. (\ref{eq:HighDensityLimit_Plus}) and (\ref{eq:MaxHubble_Minus}). Since there is no explicit time dependence in the system, one of the eigenvalues is always trivially zero, while the other two determine the stability of the solution for general power-law $(\pm)$-type models,
\begin{align}
    \lambda_{1}^{(\pm)} &= -\frac{3H}{2}\left( 1 - \frac{4 \sqrt{2}}{3} \left[\frac{(n-1) \left(\frac{25n}{32} -1 \right) \pm \left(\frac{M^2}{12H^2}\right)^{n}}{n(n-1)}\right]^{\frac{1}{2}} \right)\, ,
    \label{eq:Eigenvalue1} \\
    \lambda_{2}^{(\pm)} &= -\frac{3H}{2} \left( 1 + \frac{4 \sqrt{2}}{3} \left[\frac{(n-1) \left(\frac{25n}{32} -1 \right) \pm \left(\frac{M^2}{12H^2}\right)^{n}}{n(n-1)}\right]^{\frac{1}{2}} \right)\, ,
     \label{eq:Eigenvalue2} \\
    \lambda_3^{(\pm)} &= 0 \, .
     \label{eq:Eigenvalue3}
\end{align}

\begin{figure}[ht!]
    \centering
    \includegraphics[width=0.485\linewidth]{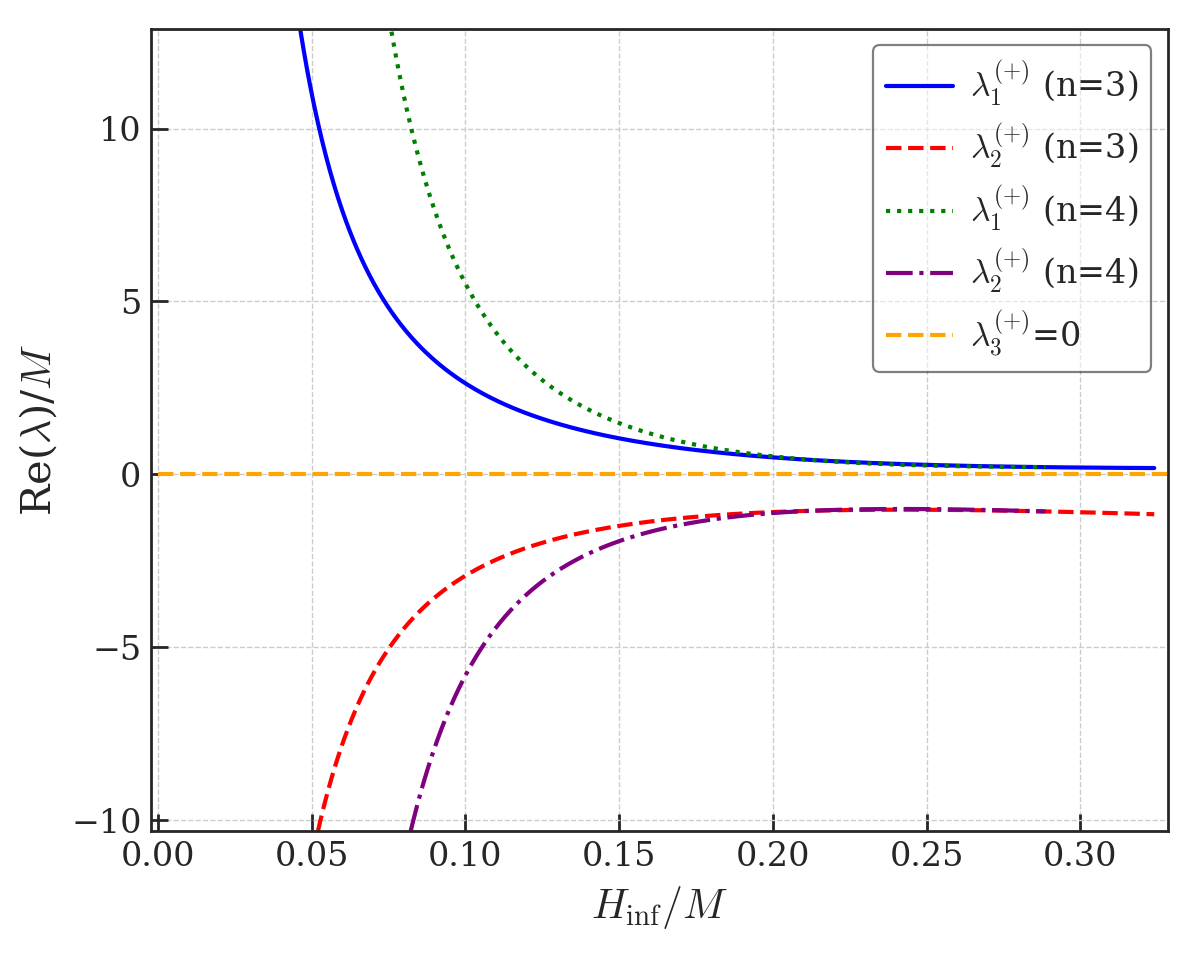}
    \includegraphics[width=0.503\linewidth]{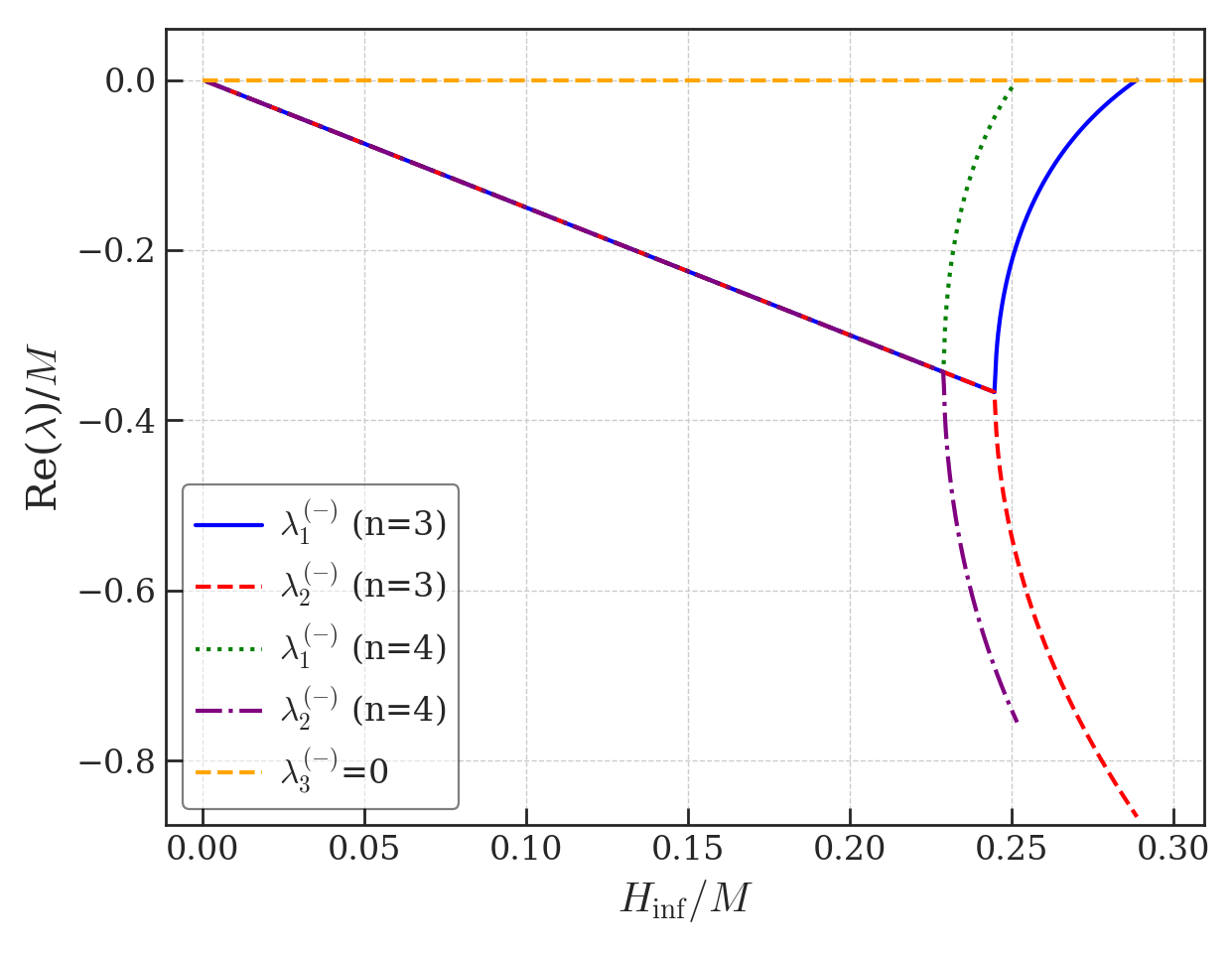}
    \caption{Eigenvalues of the dynamical system (\ref{eq:system}) for the slow-roll solution $\dot H=0$ in the $(+)$-type (left) and $(-)$-type (right) power-law NMC models up to the corresponding maximum values of $H_{\text{inf}}$.}
    \label{fig:Eigenvalues}
\end{figure}

It is evident from Eq. (\ref{eq:Eigenvalue1}) that $(+)$-type models always have one eigenvalue with a positive real part for $n\geq2$ for all $H_{\rm inf}$'s, as shown on the left side of Fig. \ref{fig:Eigenvalues}. This means that such power-law models cannot accommodate slow-roll as an attractor and imply that the GR limit is unattainable. This is in contradiction to the assumption made in Ref. \cite{Gomes:2016cwj}, where exponential expansion was taken as the definitive solution of the modified Friedmann equation (\ref{eq:GeneralFriedmannEq}), since at first glance it does satisfy the correct differential equation (\ref{eq:SlowRollDifferentialEq}). However, this is a necessary but not sufficient condition, and stable slow-roll is impossible to maintain in $(+)$-type models. There is a clear analogy for this behaviour with Dolgov-Kawasaki instabilities \cite{Dolgov:2003px} in higher-order theories, which follows from associating an effective mass to the additional degree of freedom in the theory. In the case of an NMC $f(R)$-theory, this amounts to the condition $m^2_{f(R)}\propto dF/dR>0$ \cite{Bertolami:2009cd}, which for $f_1=R$ imposes that $F_{2,R}\Lagr_m>0$. During slow-roll, $\Lagr_m<0$ and thus the requirement reduces to $F_{2,R}<0$, which is only satisfied in $(-)$-type theories. In these models, the upper bound on the inflationary scale, beyond which the real part of the first eigenvalue turns positive, is given by Eq. (\ref{eq:MaxHubble_Minus}). For Hubble scales below this threshold, it is guaranteed that $\operatorname{Re}\left[\lambda^{(-)}_{1,2}\right]<0$, which means that these models exhibit exponential expansion akin to slow-roll as attractor solutions, and can thus accommodate inflation throughout the allowed parameter space. The right side of Fig. \ref{fig:Eigenvalues} shows how $(-)$-type models satisfy the stability conditions of the $\dot H=0$ solution. In most cases, while the real parts of both eigenvalues are negative, they are also degenerate for lower values of $H_{\rm inf}$ due to the same first terms in the eigenvalues shown above. 

\subsection{Quantum fluctuations and inflationary observables}\label{subsec:QuantumFluct_CMBObservables}

Although the inflationary epoch is currently inaccessible to us via direct measurements, its echoes offer an indirect glimpse into the primordial universe, as quantum fluctuations of the inflaton field generate observable imprints on the CMB power spectrum. To determine the implications of the NMC framework on the observables resulting from the inflaton fluctuations, we consider the equation of motion for a small perturbation of the field $\delta \phi$ around the classical background $\bar \phi$ according to Eq. (\ref{eq:ModifiedConservationEq}),
where we can neglect the effective scalar field mass $V''(\bar\phi)$ and the NMC friction term for the fluctuations, even when $\Lagr_m=p$, as they occur during slow-roll where $\dot R\approx0$. Since we consider a constant curvature background, the NMC friction is negligible and the perturbations obey the same equation as in GR, such that we may use the standard result for the power spectrum in the Bunch-Davies vacuum, where the inflaton and the comoving curvature perturbations retain an identical relation during slow-roll. Thus, the dimensionless power spectrum is given in the NMC setting by 
\begin{equation}
    \Delta^2_\mathcal{R}(k)=\left(\frac{H}{2\pi}\right)^2\left(\frac{H}{\dot\phi}\right)^2= \frac{V}{24\pi^2}\left(\frac{3H^2}{V}\right)^3\epsilon_\phi^{-1} \, ,
    \label{eq:Curvature_power_spectrum}
\end{equation}
where we have used the slow-roll approximation $\dot\phi\approx-V'/3H$ and the field slow-roll parameter, which is defined as $\epsilon_\phi= \frac{1}{2} \left(\frac{V'}{V}\right)^2$. The associated scalar spectral index $n_s$ is given by
\begin{align}
    n_s-1\equiv\frac{d\ln{\Delta^2_\mathcal{R}}}{d\ln{k}}  \approx -2\frac{dH}{d\rho}\frac{V'^2}{H^3}+\frac{2}{3}\frac{V''}{H^2}=-6\frac{d(3H^2)}{d\rho}\left(\frac{V}{3H^2}\right)^2\epsilon_\phi + \frac{2V}{3H^2}\eta_\phi \, ,
\end{align}
where we have used the second slow-roll parameter $\eta_\phi= \frac{V''}{V}$. Note that the quantity $\frac{d(3H^2)}{d\rho}$ must be calculated from the modified Friedmann equation in slow-roll (\ref{eq:SlowRollFriedman}), in contrast to GR, where it is simply unity. As discussed in Ref. \cite{BarrosoVarela:2024egg}, the presence of $f_2(R)$ in the action introduces no significant modifications at leading order in the tensor sector, such that we recover the GR tensor perturbation spectrum
\begin{equation}
\Delta^2_t(k)=\frac{2H^2}{\pi^2} \, ,
\end{equation}
which is only indirectly affected by the NMC setting through the modified Hubble rate. The tensor spectral index $n_t$ is thus given by
\begin{align}\label{eq:nt}
    n_t\equiv\frac{d\ln{\Delta^2_t}}{d\ln{k}}\approx-\frac{2V'^2}{3H^3}\frac{dH}{d\rho}=-2\frac{d(3H^2)}{d\rho}\left(\frac{V}{3H^2}\right)^2\epsilon_\phi \, ,
\end{align}
while the tensor-to-scalar ratio reads 
\begin{equation}\label{eq:r}
    r\equiv\frac{\Delta^2_t}{\Delta_\mathcal{R}^2}=\frac{8}{9}\frac{V'^2}{H^4} = 16 \epsilon_{\phi} \left( \frac{V}{3H^2} \right)^2 \, .
\end{equation}
Note that using this definition we may rewrite the dimensionless curvature power spectrum as 
\begin{equation}\label{eq:r_DeltaR_relation}
\Delta_\mathcal{R}^2=\frac{2H^2}{\pi^2r} \, .
\end{equation}

These observables can be compared with constraints from the combination of 2018 Planck and BICEP/Keck (Planck-LB-BK18) results \cite{Planck:2018jri,Tristram:2021tvh},
\begin{equation}\label{eq:Planck_ns_r}
    n_s=0.9649\pm0.0042  \quad ,  \quad \quad \quad r<0.032 \ \ (95\% \ \text{CL}) \, .
\end{equation}
More recently, results from the Atacama Cosmology Telescope (ACT) combined with the aforementioned data (P-ACT-LB-BK18) \cite{ACT:2025tim} point to
\begin{equation}\label{eq:ACT_Planck_ns_r}
    n_s=0.9743\pm0.0034  \quad , \quad \quad \quad r<0.038 \ \ (95\% \ \text{CL}) \, ,
\end{equation}
which visibly shifts the constraints on the $n_s-r$ plane to higher values of $n_s$, thus compromising models so far thought to be in good agreement with the data and bringing seemingly disfavoured models back into the discussion. Due to the recent nature of these results and for the sake of easier comparison with previous works, we compare our results with both observational constraints. Note that these are significantly different from those used in Ref. \cite{Gomes:2016cwj}, where the authors used the Planck 2015 data. This is particularly important in what concerns $n_s$, which was, at the time, constrained around $1\sigma$ lower than the Planck-LB-BK18 value quoted above. In practice, one needs to numerically evolve the modified Friedmann equation given by Eq. (\ref{eq:GeneralFriedmannEq}) until inflation is over and then determine the observable quantities from the field and Hubble parameter values at 50-60 $e$-foldings from this point. An analogous analysis was conducted in Ref. \cite{Gomes:2016cwj} using a fully analytical method in dS, in contrast to the fully numerical computation beyond slow-roll in this work. Additionally, Ref. \cite{Gomes:2016cwj} focused exclusively on $(+)$-type models, which as we have shown in Sec. \ref{subsec:SlowRoll_NMC}, yield considerably different values for the viable inflationary scales and more importantly, cannot fully accommodate slow-roll expansion. 

The amplitude of the comoving curvature perturbations is determined by the CMB spectra \cite{Planck:2018jri} at the pivot scale $k_*=0.05$ Mpc$^{-1}$ to be 
\begin{equation}
    A_s\equiv\Delta_{\mathcal R}^2(k_*) \approx \left(2.10\pm0.03\right) \times 10^{-9} \, .
\end{equation}
Combining this with Eq. (\ref{eq:r_DeltaR_relation}), we can constrain the inflationary expansion rate as 
\begin{equation}
    H_{*}^2=\frac{\pi^2}{2}rA_s=\left(\frac{r}{0.01}\right)\times1.04\times10^{-10} \approx \left(\frac{r}{0.01}\right)\left(2.48\times 10^{13} \; \mathrm{GeV}\right)^2 \,,
    \label{eq:Hubble_scale_inf}
\end{equation}
for typical values of $r$ around 0.01. For $(-)$-type NMC models, the maximum physical inflationary scale given by Eq. (\ref{eq:MaxHubble_Minus}) should not be significantly below the bound (\ref{eq:Hubble_scale_inf}) to guarantee large enough expansion rates to satisfy the observational constraints. This is equivalent to a lower bound on the NMC coupling, 
\begin{equation}\label{eq:ObservationalLowerBound_M}
    M
    \gtrsim\left(\frac{(n-1)(n-2)}{2}\right)^{1/2n} \left(\frac{r}{0.01} \right)^{\frac{1}{2}} \times3.5\times10^{-5}\approx\left(\frac{r}{0.01} \right)^{\frac{1}{2}} \times8.6\times10^{13} \; \text{GeV} \, ,
\end{equation}
where in the last step we have restored the reduced Planck mass and assumed $n=3$. This places the minimum NMC scale $M$ at typical inflationary energy scales \cite{Planck:2018jri}, meaning that stronger couplings that would lead to significant post-inflationary NMC effects are ruled out. Additionally, when considering Eq. (\ref{eq:H(rho)}), this allows us to constrain the scale of the plateau of the potential as
\begin{equation}
    \begin{aligned}\label{eq:PivotScalePotential}
    V_0 \approx \rho^{(-)}_{\rm inf}
    \approx \frac{3.12\times10^{-10}}{1 +\frac{n - 2}{2}\left[ \left(\frac{r}{0.01} \right) \frac{1.25\times10^{-9}}{M^2}\right]^n} \left(\frac{r}{0.01}\right) &\gtrsim  \left(\frac{r}{0.01}\right)\frac{\left(1.03\times 10^{16} \; \mathrm{GeV}\right)^4}{n} \, , 
\end{aligned}
\end{equation}
where Eq. (\ref{eq:ObservationalLowerBound_M}) allows the inflationary scale in NMC models with $n\gtrsim3$ to be as low as $1/3$ of the equivalent prediction in GR .

\begin{figure}[ht!]
    \centering
    \includegraphics[width=0.7\linewidth]{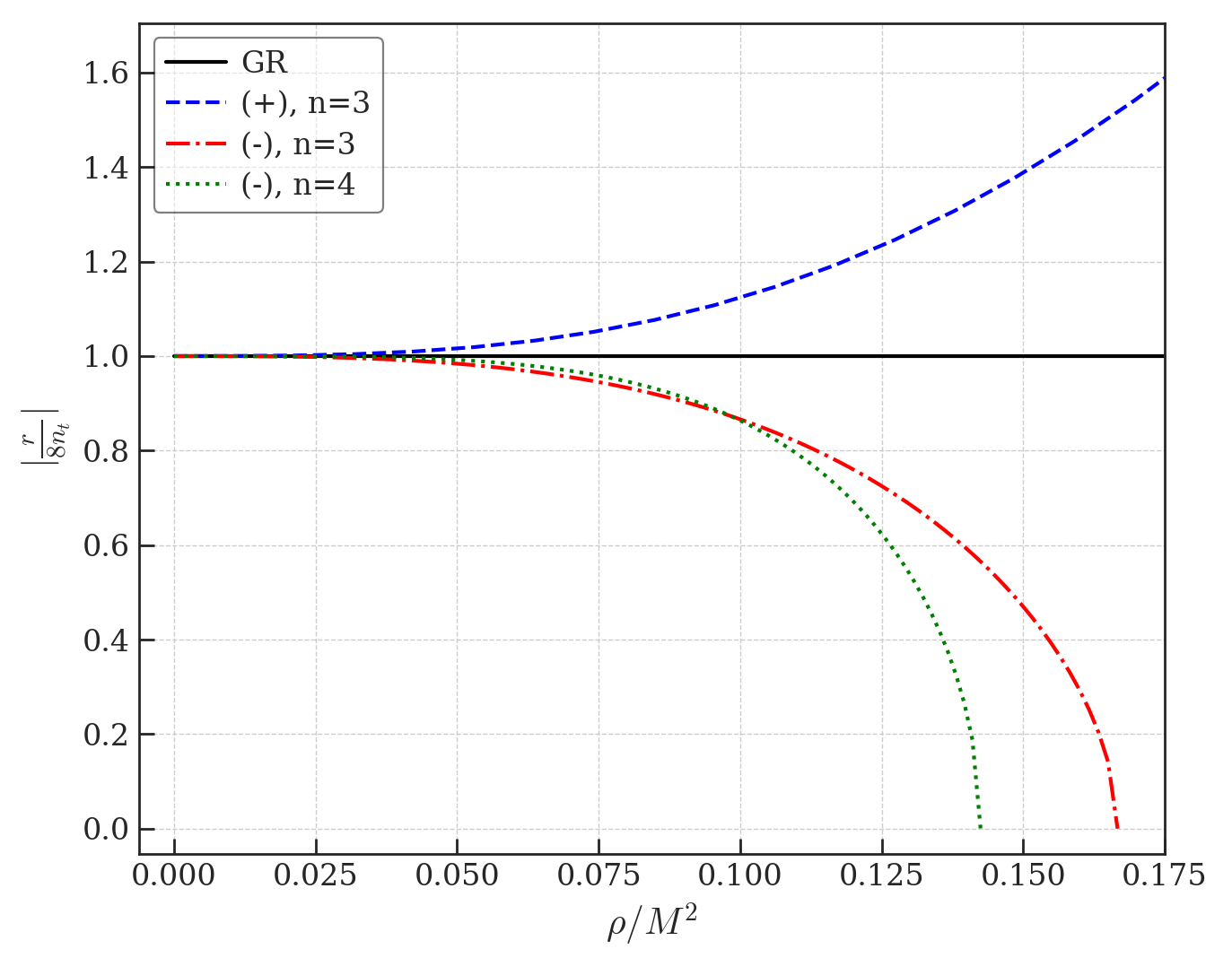}
    \caption{The ratio between the two spectral indices in $(\pm)$-type NMC models as a function of the density.}
    \label{fig:TensorRatio}
\end{figure}

Another distinctive consequence of Eqs. (\ref{eq:nt}) and (\ref{eq:r}) arises in the ratio between the two spectral indices and its deviation from the GR value due to the modified Friedmann equation for strong NMC couplings,
\begin{equation}           
    \frac{r}{8n_t}\approx-\left(\frac{d(3H^2)}{d\rho}\right)^{-1} {= \frac{4\pm(n-2)(n-1)\left( \frac{12 H^2}{M^2}\right)^n}{\left[2\mp(n-2)\left( \frac{12 H^2}{M^2}\right)^n\right]^2}} \, , 
\end{equation}
for $(\pm)$-type models respectively. In Figure \ref{fig:TensorRatio}, we see that for strong NMC couplings or large densities (in Planck units), the NMC models can deviate significantly from GR. For $n\geq2$, the $(-)$-type NMC theories start by overlapping with GR at low densities, followed by a decrease of the ratio approaching 0 as the density reaches its maximum value. 
This is considerably different from what was observed in Ref. \cite{Gomes:2016cwj} for $(+)$-type models, where this ratio initially decreased but then rapidly increased for higher density values. However, we should note that this initial decrease was a direct consequence of the approximation adopted in that work regarding the Hubble rate for $n=3$,
\begin{equation}
    H^2_{\rm inf}= \frac{2^{1/3}M^2}{12}\frac{-2^{2/3} \tilde\rho + \left( \tilde\rho^3 + \sqrt{\tilde\rho^3(4 + \tilde\rho^3)} \right)^{2/3}}{2^{1/3} \tilde\rho \left( \tilde\rho^3 + \sqrt{\tilde\rho^3(4 + \tilde\rho^3)} \right)^{1/3}}\approx\frac{2^{1/3}M^2}{12}\frac{\tilde\rho+\tilde\rho^2} {3+2 \tilde{\rho} +\tilde\rho^2} \quad \mathrm{with} \quad \tilde\rho\equiv\frac{12\rho}{2^{1/3}M^2} \, .
\end{equation}
This approximation had negligible effects for most density scales but deviates at the $\sim3\%$ level in the range $\mathcal{O}(0.01)\lesssim\rho/M^2\lesssim\mathcal{O}(1)$, leading to the emergence of that initial but unphysical decrease of the ratio. With this nuance in mind and for the sake of completeness, we include the correct prediction in Figure \ref{fig:TensorRatio}.

\subsection{Numerical evolution in the NMC model}\label{sec:NumericalEvolution}
In a full analysis of the evolution of the inflationary quantities, we must account for the coupling between the differential equations for the scalar curvature $R$ and the inflaton field $\phi$. This can be done analytically in slow-roll by disregarding the $\dot R$ terms, whereas the dynamics have to be computed numerically beyond slow-roll. These coupled equations are higher-order than GR and therefore can be quite unstable in their numerical evaluation. We can obtain the full dynamic behaviour of spacetime curvature through the general modified Friedmann Eq. (\ref{eq:GeneralFriedmannEq}) with $\Lagr_m$ from Eq. (\ref{eq:Lagr_m}), and using Eqns. (\ref{eq:F}) and (\ref{eq:ModifiedConservationEq}),
\begin{equation}\label{eq:dRdN}
    \frac{dR}{dN}=f_2\frac{-f_1 + 24\beta\dot\phi V'F_2H - 6FH^2 + \dot\phi^2\left(f_2 +36 (2\beta-1)F_2H^2\right) + 
     FR + 2f_2V}{6H^2 \left[ 2(1-2\beta)\dot\phi^2F_2^2 + F_{2,R}f_2\left((2\beta-1)\dot\phi^2 - 2V\right) \right]} \, ,
\end{equation}
where we have chosen $e$-folds defined as $dN=Hdt$ as our numerical ``time'' coordinate. Note that Eq. (\ref{eq:dRdN}) has a non-trivial GR limit when $f_2\rightarrow1$, since both the numerator and the denominator vanish by definition, meaning that we must be careful when evolving the system of equations to ensure a smooth transition to the weakly coupled regime. The evolution of the Hubble rate follows from $R=6H(2H+\frac{dH}{dN})$, and similarly for the inflaton field value via $\dot\phi=H\frac{d\phi}{dN}$, where the inflaton velocity obeys the modified conservation equation (\ref{eq:ModifiedConservationEq}). By evolving these quantities simultaneously, we ensure an accurate and consistent evaluation of the inflationary dynamics of the NMC model without any assumptions apart from the initial slow-roll boundary conditions. Although we have 4 quantities to evolve in the first-order differential equations for the system $\{\phi,\dot\phi,H, R\}$, they are not independent from each other. The slow-roll condition at the start of the numerical evolution dictates that we only need 2 initial conditions, which can be given in terms of $\{\phi_{\rm inf},\dot\phi_{\rm inf}\}$, as the density depends on the initial field conditions via $\rho_{\rm inf}=\dot\phi_{\rm inf}^2/2+V(\phi_{\rm inf})$, the Hubble rate scales with the density according to Eq. (\ref{eq:H(rho)}), and $R_{\rm inf}=12H_{\rm inf}^2$.

The parameter space of the theory is determined by the inflationary potential $V(\phi)$, which will typically range from 1 to 3 parameters, along with the NMC parameter $M$ in $f_2$. In total, we have 2 initial conditions and 2-4 theory parameters, which can be constrained by the Planck measurements \cite{Planck:2018jri}. In general models of cold inflation, we can typically take $\dot\phi_*$ to be arbitrarily close to zero during inflation, which simplifies the initial conditions. In order to determine the inflationary observables described in Section \ref{subsec:QuantumFluct_CMBObservables}, one should look at 50-60 $e$-folds before the end of inflation, which requires establishing the conditions for this halting point. To that end, we use the first slow-roll parameter $\epsilon=-\dot H/H^2$, which may be determined in terms of the numerically estimated dynamical variables, $\epsilon=2-\frac{R}{6H^2}$. We start in slow-roll where $\epsilon_{\rm inf}=0$, the universe undergoes inflationary expansion while $\epsilon<1$ and the potential dominates the kinetic energy, $\dot\phi^2/2<V$. Once one of these conditions is no longer satisfied, we can consider inflation to be over and determine the CMB observables from 50-60 $e$-foldings before that point. Note that this is different from Ref. \cite{Gomes:2016cwj}, where the end of inflation was determined analytically using the violation of the field slow-roll parameter $\epsilon_\phi$ when crossing the critical value $\epsilon_{\phi,\text{end}}=3\left(\frac{H^2}{V}\right)^2\frac{d\rho}{dH^2}$. This approach assumes that all quantities can be expressed in terms of their slow-roll values, e.g. $\rho\sim V$, which becomes an increasingly inaccurate approximation as we approach the latter stages of inflation. In fact, in our full numerical analysis of the inflationary dynamics both in GR and in the NMC model, this threshold is often not crossed, although $\epsilon=1$ is achieved. Hence, the analytical estimates of the end of inflation in Ref. \cite{Gomes:2016cwj} underestimate the end of inflation by up to 50$\%$ for the NMC model.

\section{Theoretical predictions for inflationary observables}\label{sec:Results}

Many of the details of the inflationary period are hidden behind the curtain of early universe dynamics, making it difficult to distinctly test particular properties of different theories. However, as discussed in Section \ref{subsec:QuantumFluct_CMBObservables}, quantum fluctuations of the inflaton field lead to observable imprints on the CMB spectrum such as $n_s$ and $r$, which allow for comparison of different gravitational theories and inflationary models. To determine the scale of the inflationary models under consideration, we solve Eq. (\ref{eq:Curvature_power_spectrum}) for the inflaton field value at the pivot scale $\phi_*$, as shown in Eq. (\ref{eq:PivotScalePotential}). Although the precise estimation of the time that the pivot scale leaves the horizon depends on the post-inflationary era, it is usually assumed that it happens 50-60 $e$-folds before the inflationary finale \cite{Mantziris:2020rzh, Liddle:2000cg}. 

In this section, we present theoretical predictions for inflationary observables from the numerical evaluation of the inflationary dynamics for various models in $(-)$-type NMC theories. We focus on $n=3$ as the lowest order that deviates from GR, since $n=2$ reproduces GR in the slow-roll regime, as discussed in Section \ref{sec:NMCInflation} and confirmed numerically in what follows. Since the Hubble scale depends only on the ratio $\rho/M^2$ according to Eq. (\ref{eq:H(rho)}), the quantity that enters the calculations is the ratio $V_0/M^2$, where $V_0$ is the scale of the slow-roll plateau constrained by Eq. (\ref{eq:PivotScalePotential}). Therefore, we quote the value of $M^2/V_0$ (in Planck units) in each plot, where it is important to keep in mind that smaller values correspond to stronger couplings. In order to probe significant deviations from GR while staying within the maximum density value of Eq. (\ref{eq:MaxRho_Minus}), we fix $M^2/V_0=9$, safely above the minimum value of $M^2/V_0\sim6$, as seen in Fig. \ref{fig:MaxHubble&Density}. With these considerations in mind, we present the comparison of the resulting parameter space in the $n_s-r$ plane for some common inflationary models between GR and the NMC theory settings.

\subsection{Polynomial potentials: quartic symmetry-breaking, hilltop, and natural inflation}

\begin{figure}[ht!]
    \centering
    \includegraphics[width=0.8\linewidth]{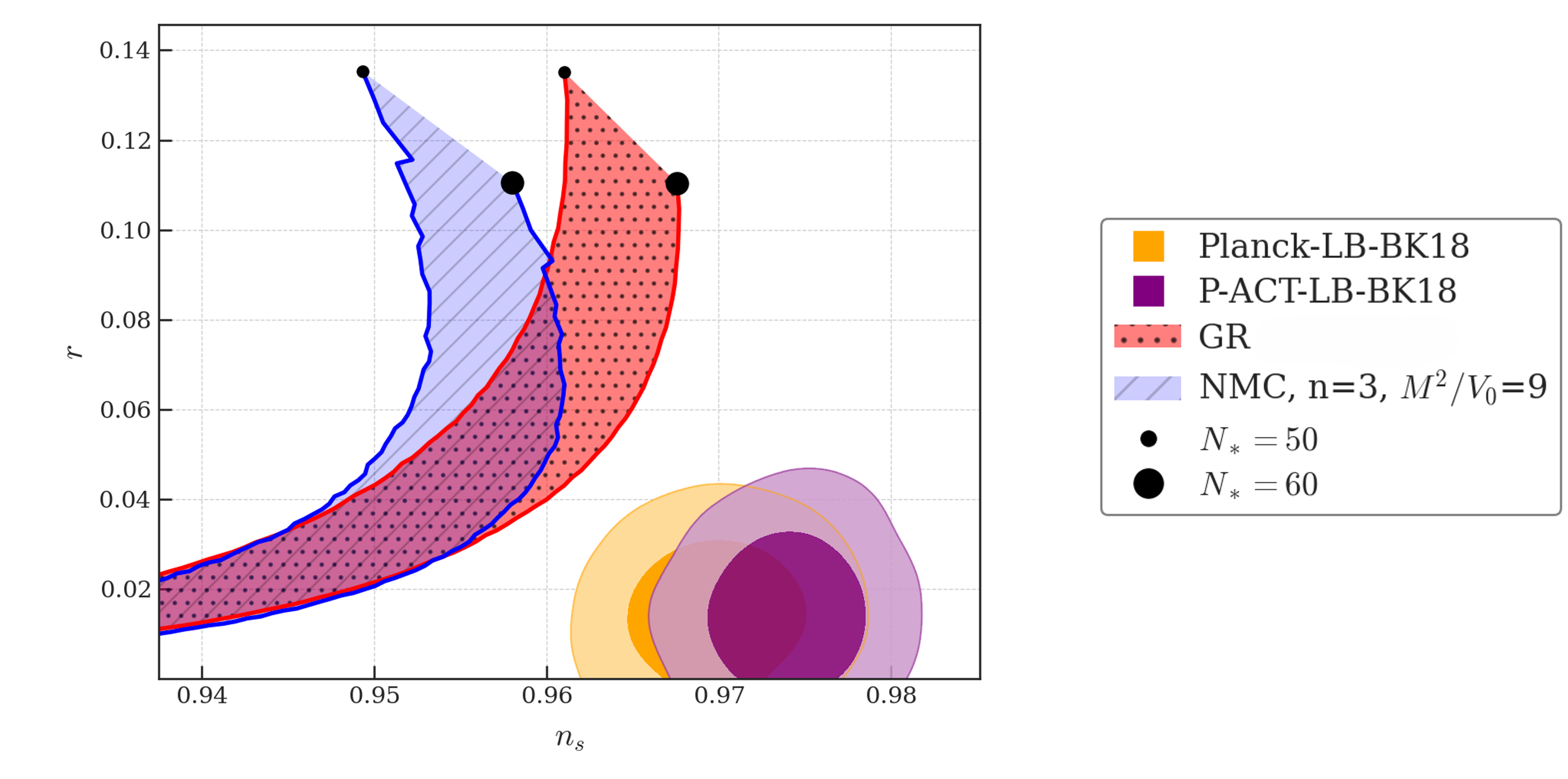}
    \caption{Predictions for the scalar-to-tensor ratio and the scalar spectral index with a quartic symmetry-breaking potential, where $\gamma$ decreases from left to right in the range $[0.05,0.001]$, and the boundaries of the shaded region correspond to inflation lasting $N_*=50-60$ $e$-folds. The $n=2$ NMC model matches GR, but both $n=2,3$ are over $2\sigma$ away from the observational measurements. }
\label{fig:QuarticSB_Posteriors}
\end{figure}

A quartic symmetry-breaking potential can be written generally as
\begin{equation}
    %V(\phi)=V_0\left[1-\frac{\gamma}{2}\phi^2+\frac{\gamma^2}{16}\phi^4\right] \, ,
    V(\phi)=V_0 \left[1-\frac{\gamma}{4}\phi^2\right]^2 \, ,
\end{equation}
where $\gamma$ regulates the width of the potential wells. The inflaton starts close to the top of the potential at $\phi\approx0$ and rolls down towards the minimum at $\phi=\frac{2}{\sqrt{\gamma}}$. For the same initial conditions, varying $\gamma$ changes the duration of inflation along with the observables $n_s$ and $r$. It is clear in Fig. \ref{fig:QuarticSB_Posteriors} that even for a strong coupling ($M^2/V_0=9$), the $n=2$ NMC model mimics GR during slow-roll. Although these potentials are ruled out well beyond $2\sigma$ in GR, we arrive at a contrasting conclusion compared to Ref. \cite{Gomes:2016cwj}, where 2015 Planck constraints exhibited a preference for lower values of $n_s$ with higher values of $r$. The range of the inflationary observables for the $(-)$-type model with $n=3$ deviate from GR towards higher values of $r$, which disfavours these models even more decisively.

\begin{figure}[ht!]
    \centering
    \includegraphics[width=\linewidth]{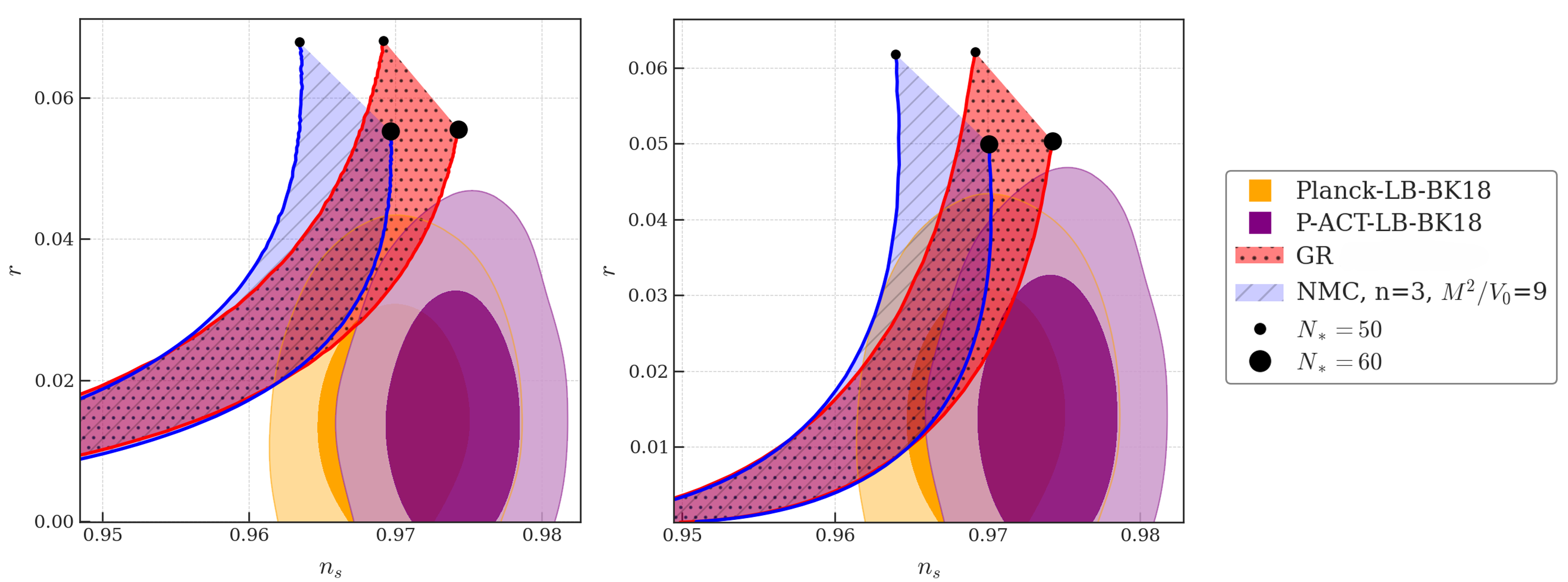}
    \caption{Predictions for the scalar-to-tensor ratio and the scalar spectral index with a quadratic (left) and a quartic (right) hilltop potential, where $\gamma$ decreases from left to right within $[1,0]$ and the shaded region corresponds to $N_*=50-60$ $e$-folds for the duration of inflation. The NMC model with $n=2$ matches GR, while $n=3$ recedes from the $2 \sigma$ (left) and $1\sigma$ (right) areas of the observational constraints. }
\label{fig:Hilltop_Posteriors}
\end{figure}

We can write a general hilltop potential as
\begin{equation}
    V(\phi)=V_0\left[1-\frac{\gamma}{m}\phi^m\right] \, ,
\end{equation}
where $\gamma$ sets the shallowness of the potential \cite{Boubekeur:2005zm}. The inflaton starts at the top of the potential at $\phi\approx0$ and rolls down until its energy is no longer dominated by the potential. The resulting inflationary observables for the quadratic hilltop potential ($m=2$) are shown in the left plot of Fig. \ref{fig:Hilltop_Posteriors}, where $n=2$ mimics GR and thus remains within the $(1-2)\sigma$ region of the observational constraints. However, note that they are not as favourable, as in Ref. \cite{Gomes:2016cwj}, when compared with the current measurements \cite{Planck:2018jri, Tristram:2021tvh, ACT:2025tim}. The $n=3$ NMC model favours larger values of the tensor-to-scalar ratio $r$, which is ruled out by the P-ACT-LB-BK18 data at $2\sigma$. The results follow a similar pattern for the quartic ($m=4$) hilltop potential, although both GR and the $n=3$ NMC model predict smaller values of $r$ and are therefore in better agreement with the observational constraints.

The behaviour of the inflaton in natural inflation \cite{Freese:1990rb} resembles that of a quadratic hilltop model during slow-roll, when the field is near the top of the potential ($\phi\ll1)$, so we include it here for comparison, 
\begin{equation}
    V(\phi)=V_0\left[1+\cos \left({\frac{\phi}{f}} \right)\right]\approx 2V_0\left[1-\frac{\phi^2}{4f^2}\right] +\mathcal{O}(\phi^4)\, .
\end{equation}
Very similar conclusions can be drawn for this family of potentials, placing them more within the constraints of Planck-LB-BK18, but being slightly disfavoured by the more recent P-ACT-LB-BK18 data.

In contrast to Ref. \cite{Gomes:2016cwj}, we do not consider chaotic models with monomial potentials described by $V\propto\phi^m$,  due to their characteristically large tensor-to-scalar ratios $r\sim\frac{4m}{N_*}$, i.e. for example $r\gtrsim0.1$ for $m=2$. The 95\% confidence level constraints place $r\lesssim0.03$ in irreparable tension with the chaotic potential predictions \cite{Planck:2018jri}. Even for the linear monomial potential ($m=1$), these constraints point to the problematic relation between chaotic models and observational data, which would only be worsened by the higher tensor-to-scalar ratios we have found to be associated in the NMC inflationary models thus far. 

\subsection{Starobinsky-like potential}

One of the most minimal and favourable models with respect to the CMB observations has been Starobinsky inflation~\cite{Starobinsky:1980te, Starobinsky:1982mr, Barrow:1983rx, Vilenkin:1985md, DiValentino:2016nni}, where the inflaton field arises from a propagating scalar degree of freedom in the action due to a quadratic curvature term in the Jordan frame \cite{Whitt:1984pd}. In the context of this work, this would translate to having $f_1=R+R^2/6M^2_1$, where the free parameter is fixed by the CMB in the Einstein frame at approximately the inflationary scale \cite{Kofman:1985aw, Maeda:1987xf, Barrow:1988xh, Barrow:1988xi}, $M_1\sim \mathcal{O} \left(10^{13}\right)$ GeV \cite{Gorbunov:2012ns, Mantziris:2020rzh, Mantziris:2022fuu}. Such higher order terms in the gravitational sector can usually be argued from an EFT perspective \cite{Donoghue:1994dn}, renormalizability \cite{Percacci:2025ehx}, and the gravitational-induced renormalization group running or the conformal anomaly ~\cite{Markkanen:2013nwa,Markkanen:2018bfx}, but the verdict regarding the swampland conjectures seems currently unfavourable \cite{Lust:2023zql, Antoniadis:2025pfa} but inconclusive \cite{Ketov:2024klm}. It is worth noting that in scenarios where considering other scalar fields (e.g. the SM Higgs) is necessary, there will be mixing due to derivative couplings between the scalar fields \cite{Kehagias:2013mya, Wang:2017fuy, Ema:2017rqn, He:2018gyf, Gorbunov:2018llf, Bezrukov:2019ylq, Ema:2020evi, Li:2022ugn}. In the presence of a non-trivial $f_2$ term, as in this study, performing the conformal transformation would exhibit some complications for a non-zero matter Lagrangian. However, since there is no inflationary matter sector in the Jordan frame in Starobinsky inflation, the transfer to the Einstein frame can be performed as usual \cite{Mantziris:2022fuu, Li:2021fao}. Therefore, as the inflaton is always decoupled from $f_2$ in Starobinsky inflation, here we study the Starobinsky-like potential,
\begin{equation}
    V(\phi)= {V_0} \left[1-e^{-\sqrt{\frac{2}{3 
    }}\phi}\right]^2 \, ,
\end{equation}
in the Einstein frame where $f_1=R$, and the inflationary Lagrangian is coupled to curvature via $f_2$.

\begin{figure}[ht!]
    \centering
    \includegraphics[width=0.94\linewidth]{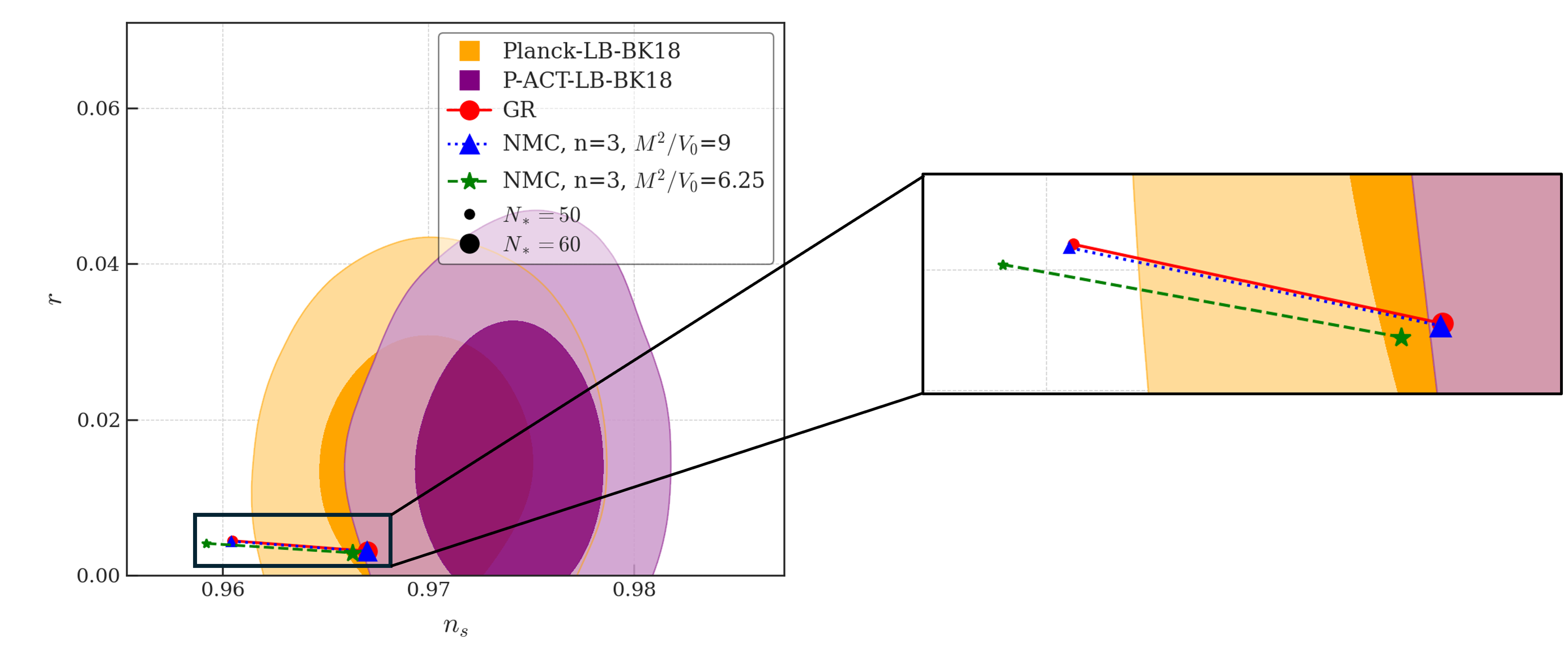}
    \caption{Predictions for the scalar-to-tensor ratio and the scalar spectral index with a Starobinsky-like potential. The $n=3$ NMC model matches the GR predictions except when considering very strong couplings, as shown by the green dashed curve with the stars.}
    \label{fig:Starobinsky_Posteriors}
\end{figure}

The resulting predictions for this potential are shown in Fig. \ref{fig:Starobinsky_Posteriors}, specifically for GR and the $n=3$ NMC model. We see that the modified theory leads to identical observables to classical gravity, even when considering a moderately strong coupling $M^2/V_0=9$, illustrating that Starobinsky-like potentials are less sensitive to the NMC theory than the other potentials considered. However, for very strong couplings near the maximum allowed value of $\rho/M^2$, such as $M^2/V_0\sim6$, there is a slight deviation of the observables from GR predictions to lower values of $r$ and $n_s$. This area of the parameter space is slightly more disfavoured, although still within the 1$\sigma$ region of Planck-LB-BK18 for longer inflationary epochs. We omit the resulting predictions for the $n=2$ NMC model for brevity, since they are identical to GR \cite{Drees:2025ngb}, as in the case of polynomial potentials above.

\subsection{Old inflation}\label{subsec:OldInflation}

\begin{figure}[ht!]
    \centering
    \includegraphics[width=0.6\linewidth]{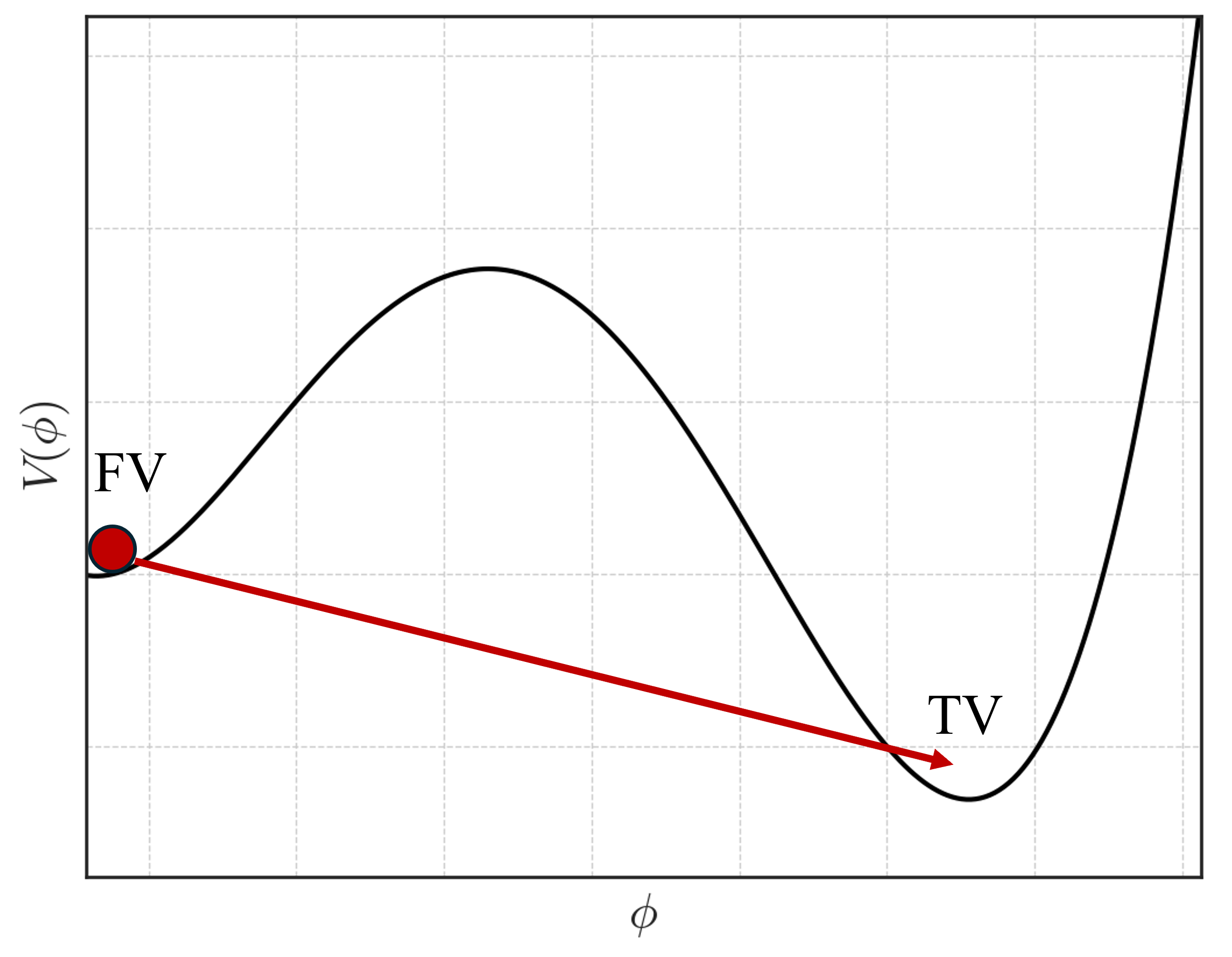}
    \caption{Double-well potential of ``old'' inflation, where the inflaton starts in the false vacuum (FV) and the universe inflates exponentially, after which the field tunnels to the true vacuum (TV) and inflation ends. }
    \label{fig:CubicPotential}
\end{figure}

Our analysis (and Ref. \cite{Gomes:2016cwj}) has focused on models of ``new'' inflation, where the inflationary era is terminated by a second-order phase transition \cite{Linde:1981mu,Albrecht:1982wi}, For completeness, we consider the effects of the NMC framework on the ``old'' inflationary scenario \cite{Guth:1980zm}, where inflation is halted due to a first-order phase transition from a false vacuum (FV) that behaved as vacuum energy, $p\approx-\rho \approx - V(\phi)$, to a sub-dominant true vacuum (TV), as illustrated in Figure \ref{fig:CubicPotential}. The inflaton tunnels through the potential barrier with a constant decay rate per spacetime volume $\Gamma$ leading to the formation of TV bubbles that expand approximately at the speed of light. However, inflation is complete only when the TV bubbles percolate, i.e. when $\varepsilon=\Gamma/H^4 > 1$. As shown in Appendix \ref{sec:OldInflationFailureProof}, in GR not only does $\varepsilon$ need to be large enough, $\varepsilon_{\rm min} \sim0.7$, to guarantee percolation\footnote{In Ref. \cite{Guth:1982pn}, more technical arguments place this lower bound between $1.1\times10^{-6} \leq \varepsilon_{\rm min} \leq 0.24$.}, but simultaneously smaller than $\varepsilon_{\rm max}\sim4\times10^{-3}$ to ensure that sufficient inflation has occurred \cite{Guth:1982pn}. Evidently, only a very finely tuned area of the parameter space can satisfy both requirements, which is known as the ``graceful exit'' problem of ``old'' inflation.

For a modified theory to solve the graceful exit problem, the ideal scenario would start with a very low $\varepsilon$ that increases over time in order to ensure both the necessary inflationary expansion and bubble percolation with no need for fine-tuning. For example, extended inflation \cite{La:1989st}, which corresponds to a scalar-tensor theory with a time-dependent effective gravitational constant, is characterised by a time-varying Hubble rate during slow-roll. This feature can accommodate a transition from an initial phase of exponential expansion to a subsequent phase of accelerated power-law expansion, which allows inflation to last for the required number of $e$-foldings and the TV bubbles to percolate after nucleation. Such an approach is only successful for low values of the Brans-Dicke parameter $\omega_{\rm BD}<25$, as described in Ref. \cite{La:1989pn}. However, observational tests constrain this parameter to significantly larger values. For example, data from a pulsar in a triple star system imposes that $\omega_{\rm BD}>1.4\times10^5$ \cite{Voisin:2020lqi}, therefore ruling out extended inflation.

Could something similar be achieved in the NMC model without the drawbacks of extended inflation? In general, NMC models can lead to effects in the effective gravitational constant, similarly as in extended inflation. Unlike Ref. \cite{La:1989za}, this effective constant, $G_{\rm eff}=\frac{f_2}{1+2F_2\Lagr_m}G$, would be fully determined for given values of $R$ and $\rho$, which are constant in slow-roll and therefore cannot lead to a dynamic Hubble rate that would evolve from exponential to power-law-like expansion. Additionally, a power-law form for $f_2$ according to Eq. (\ref{eq:f2}) implies that NMC effects get weaker with time as $R$ decreases, so it is impossible to establish a transition from exponential to power-law inflation as in Ref. \cite{La:1989za}. Given that inflationary expansion is still exponential in the NMC context, the argument presented in Appendix \ref{sec:OldInflationFailureProof} naturally applies to this model for $\Gamma \approx$ constant, regardless of the strength of the NMC coupling $M$. Moreover, the high-density regime of $(-)$-type NMC theories naturally provides larger values of $H_{\rm inf}$ that lead to smaller $\varepsilon$' s, which further magnifies the graceful exit problem.
Thus, we conclude that ``old'' inflation cannot be rescued solely by the inclusion of a $f_2(R)$ non-minimal coupling between matter and curvature.

\section{Conclusions}\label{sec:Conclusions}
In this work, we have examined the implications of NMC $f(R)$-theories on the background and field dynamics during inflation. After introducing NMC matter-curvature models from a cosmological perspective, we extended the analysis of Ref. \cite{Gomes:2016cwj} to include negative higher-order corrections to the matter-curvature coupling. We generalised the treatment to both $(\pm)$-type models and studied their effects on the dS phase, finding clear distinctions in the stability and compatibility of the allowed slow-roll solutions via the eigenvalues of the dynamical system's Jacobian matrix. We found that during slow roll, the dS solution in $(-)$-type models is always a stable attractor for the entire allowed parameter space, and possesses a maximum physical limit on both the density and the Hubble rate subject to the NMC scale $M$. In contrast, $(+)$-type models always have unstable slow-roll solutions, making exponential expansion impossible to maintain within this kind of NMC models.

We provided an overview of the emergence of observational imprints in the CMB spectrum from inflationary fluctuations and the current state-of-the-art measurements, and showcased the corresponding consequences from NMC models. More specifically, we showed how experimental measurements constrain the strength of the NMC coupling around typical inflationary energy scales, $M_{\rm min}\sim10^{13}-10^{14}$ GeV, as expected. This firmly places this kind of GR modification on scales close and above the inflationary ones, as lower scales would lead to tight limits on the inflationary Hubble rate, thus directly contradicting observations for the amplitude of the scalar power spectrum at the pivot scale.

We developed a general and robust numerical method that evolves the system from the initial conditions in dS until the end of inflation, determined by the end of exponential expansion according to the first slow-roll parameter $\epsilon$. This allowed us to analyse various inflationary potentials and their corresponding observational predictions in GR and within a family of NMC models. For both the quartic symmetry-breaking and hilltop potentials, we found that the modified theory points to higher values of the tensor-to-scalar ratio $r$, leading to tension with the most recent constraints on the $n_s-r$ plane when considering strong non-minimal couplings. Although ``pure'' Starobinsky inflation is decoupled from such NMC theories due to the generation of the inflaton field in the Einstein frame from the conformal transformation of the $R^2$ term, we considered a Starobinsky-like potential and found that for couplings weaker than $M^2/V_0\sim9$, the NMC setting has little to no effect on the inflationary observables, placing it safely within the Planck-LB-BK18 bounds but just outside the $1\sigma$ region of P-ACT-LB-BK18. Our results point to tight bounds on the non-minimal coupling between matter and curvature, which for most of the considered potentials practically restricts the modifications to the minimal coupling in the GR action to be slightly over the perturbative level at best. Although GR is favoured by the data, there are viable areas of the parameter space that can support NMC inflationary models that can be distinguished from classical gravity. We have also shown that this modified theory is unable to salvage ``old'' inflation from the graceful exit problem, as exponential expansion inevitably hinders this model with its inability for bubble percolation after sufficient dS expansion.

The numerical computation of the inflationary dynamics of the NMC theories was built with minimal assumptions without the slow-roll approximation, allowing us to derive insightful conclusions and paving the way for extending this analysis to the post-inflationary epoch. An interesting and natural extension of this work would extend our framework to (p)reheating scenarios \cite{Bertolami:2010ke}, where not only the friction term in Eq. (\ref{eq:ModifiedConservationEq}) is important, but the introduction of direct decays of the inflaton to (B)SM particles, scatterings, or gravity-mediated particle production can lead to other non-trivial matter-curvature interactions \cite{Barman:2025lvk}. In fact, for ``new'' inflation models that include a prolonged period of kination before reheating, there can be sizeable values of $\Gamma_c\propto\dot R$ that could temporarily overcome the Hubble friction, depending on the rate at which the transition from inflation to kination occurs. In the interest of preserving the focus of this paper on the implications of the NMC theory on the inflationary era, we postpone this study for our next endeavour.

\begin{acknowledgments}
    The work of M.B.V. was supported by FCT through the grant 2024.00457.BD. The work of A.M. and O.B. (partially) was supported by FCT - Fundação para a Ciência e Tecnologia, I.P. through the project with DOI identifier 10.54499/2024.00252.CERN. 
\end{acknowledgments}

\appendix

\section{Graceful exit in old inflation} \label{sec:OldInflationFailureProof} 

In this section, we provide a brief overview of the argument from Ref. \cite{Guth:1982pn} regarding the shortcoming of ``old'' inflation in achieving percolation and thus solving the graceful exit problem \cite{Liddle:2000cg}. The probability that a point is found in the FV is given by \cite{Guth:1982pn} to be
\begin{equation}
    P_{\rm fv}(t)=\exp\left[-\int_{t_{\rm nuc}}^t dt' \Gamma(t')a^3(t')\mathcal{V}(t,t')  \right] \,,
\end{equation}
where $\mathcal{V}(t,t')$ is the volume of a bubble at time $t$ which nucleated at time $t'$, and is given by 
\begin{equation}
    \mathcal{V}(t,t')=\frac{4}{3}\pi\left[\int dr\right]^3=\frac{4}{3}\pi\left[\int_{t'}^t \frac{dt''}{a(t'')}\right]^3 \, ,
\end{equation}
where $r$ is the bubble radius and its expansion velocity is approximately equal to the speed of light. For exponential expansion $a(t)=e^{H_{\rm inf} t}$ and a time-independent transition rate $\Gamma$, we get the probability
\begin{equation}
    P_{\rm fv}(t)\sim\exp\left[-\frac{4\pi \Gamma}{3H_{\rm inf}^3} (t-t_{\rm nuc})\right]=\exp\left[-\frac{4\pi \varepsilon H_{\rm inf}}{3} (t-t_{\rm nuc})\right] \,,
\end{equation}
which decreases exponentially with time. However, the volume of the universe still in the FV scales as $ \mathcal{V}_{\rm fv}(t)=a^3(t)P_{\rm fv}(t)$ and evolves for $t\gg t_{\rm nuc}$ as
\begin{equation}
    V_{\rm fv}\sim \exp\left[(3H_{\rm inf}-\frac{4\pi}{3}\varepsilon H_{\rm inf})t\right] \,,
\end{equation}
which decreases exponentially with time only if the condition $3H_{\rm inf}<\frac{4\pi}{3}\varepsilon H_{\rm inf}$ is satisfied for a given inflationary model. Therefore, bubble percolation is achieved only for $\Gamma/H^4_{\rm inf}>\varepsilon_{\rm min} =9/4\pi\approx0.71$. Considering that between the start of the inflationary phase transition and the time of percolation, the universe must expand for $N_*=50-60$ $e$-folds by a factor of $e^{N_*}$, we can estimate the maximum allowed value of $\varepsilon$, since $P_{\rm fv}$ decays exponentially with a time constant $\tau^{-1}=\frac{4\pi}{3}\varepsilon H_{\rm inf}$,
\begin{equation}
    e^{N_*}=e^{H_{\rm inf}\tau}\Rightarrow\varepsilon<\varepsilon_{\rm max} =\frac{3}{4\pi N_*}\sim4\times10^{-3} \, .
\end{equation}
Since $\varepsilon_{\rm max}$ is smaller than $\varepsilon_{\rm min}$, it is impossible to have enough inflation and a graceful exit in an inflationary model with a time-independent $\Gamma$ and fully exponential inflation.  

\bibliographystyle{JHEP}
\bibliography{References.bib}

\end{document}